\documentclass[conference,twocolumn,letterpaper,10pt]{IEEEtran}

\pdfoutput=1
\usepackage[cmex10]{amsmath}
\usepackage{amsfonts}
\usepackage{verbatim}

\usepackage{cite}
\ifCLASSINFOpdf
  \usepackage[pdftex]{graphicx}
  \graphicspath{{graphics/}}
  \DeclareGraphicsExtensions{.pdf,.jpeg,.png}
\else
  \usepackage[dvips]{graphicx}
  \graphicspath{{graphics/}}
  \DeclareGraphicsExtensions{.eps}
\fi

\usepackage{array}
\usepackage{mdwmath}
\usepackage{mdwtab}
\DeclareMathOperator{\erf}{erf}

\begin{document}
\bibliographystyle{IEEEtran}
\title{Using Dimensional Analysis to Assess Scalability and Accuracy in
Molecular Communication}

\author{\IEEEauthorblockN{Adam Noel$^{\star\dagger}$, Karen C.
Cheung$^{\star}$, and Robert Schober$^{\star\dagger}$}
\IEEEauthorblockA{$^{\star}$Department of Electrical and Computer
Engineering\\
University of British Columbia, Email: \{adamn, kcheung, rschober\}@ece.ubc.ca
\\ $^{\dagger}$Institute for Digital Communications\\
University of Erlangen-N$\ddot{\textnormal{u}}$rnberg, Email: \{noel,
schober\}@LNT.de}}


% Calculus
\newcommand{\dbydt}[1]{\frac{d#1}{dt}}
\newcommand{\pbypx}[2]{\frac{\partial #1}{\partial #2}}
\newcommand{\psbypxs}[2]{\frac{\partial^2 #1}{\partial {#2}^2}}
\newcommand{\dbydtc}[1]{\dbydt{\conc{#1}}}
\newcommand{\thev}{\theta_v}
\newcommand{\thevi}[1]{\theta_{v#1}}
\newcommand{\theh}{\theta_h}
\newcommand{\thehi}[1]{\theta_{h#1}}
\newcommand{\x}{x}
\newcommand{\y}{y}
\newcommand{\z}{z}
\newcommand{\rad}[1]{\vec{r}_{#1}}
\newcommand{\radmag}[1]{|\rad{#1}|}

% Chemical parameters
\newcommand{\kth}[1]{k_{#1}}
\newcommand{\km}{K_M}
\newcommand{\vm}{v_{max}}
\newcommand{\conc}[1]{[#1]}
\newcommand{\conco}[1]{[#1]_0}
\newcommand{\C}{C}
\newcommand{\Cx}[1]{C_{#1}}
\newcommand{\CxFun}[3]{C_{#1}(#2,#3)}
\newcommand{\Cobs}{C_{obs}}
\newcommand{\Nobs}{{\Nx{\A}}_{obs}}
\newcommand{\Nobst}[1]{\Nobs\left(#1\right)}
\newcommand{\Nobsn}[1]{\Nobs\left[#1\right]}
\newcommand{\Nobsavgt}{\overline{{\Nx{\A}}_{obs}}(t)}
\newcommand{\Nobsavg}[1]{\overline{{\Nx{\A}}_{obs}}\left(#1\right)}
\newcommand{\Nobsavgmax}{\overline{{\Nx{\A}}_{max}}}
\newcommand{\Cgen}{C_A(r, t)}
\newcommand{\radbind}{r_B}

% Physical parameters
\newcommand{\M}{M}
\newcommand{\smM}{m}
\newcommand{\A}{A}
\newcommand{\X}{S}
\newcommand{\metre}{\textnormal{m}}
\newcommand{\second}{\textnormal{s}}
\newcommand{\molecule}{\textnormal{molecule}}
\newcommand{\bound}{\textnormal{bound}}
\newcommand{\Dx}[1]{D_{#1}}
\newcommand{\Nx}[1]{N_{#1}}
\newcommand{\Da}{D_\A}
\newcommand{\En}{E}
\newcommand{\en}{e}
\newcommand{\Ne}{N_{\En}}
\newcommand{\De}{D_\En}
\newcommand{\EA}{EA}
\newcommand{\ea}{ea}
\newcommand{\Nint}{N_{\EA}}
\newcommand{\Di}{D_{\EA}}
\newcommand{\Etot}{\En_{Tot}}
\newcommand{\stepl}{r_{rms}}
\newcommand{\AP}{A_P}
\newcommand{\Ri}[1]{R_{#1}}
\newcommand{\ro}{r_0}
\newcommand{\rone}{r_1}
\newcommand{\visc}{\eta}
\newcommand{\bolt}{\kth{B}}
\newcommand{\temp}{T}
\newcommand{\T}{T_B}
\newcommand{\Vobs}{V_{obs}}
\newcommand{\robs}{r_{obs}}
\newcommand{\Ve}{V_{enz}}
\newcommand{\tint}{\delta t}
\newcommand{\tmax}{t_{max}}
\newcommand{\Cobsfrac}{\alpha}
\newcommand{\dist}{L}
\newcommand{\DMLSA}{a}
\newcommand{\DMLSt}[1]{t_{#1}^\star}
\newcommand{\DMLSx}{x^\star}
\newcommand{\DMLSy}{y^\star}
\newcommand{\DMLSz}{z^\star}
\newcommand{\DMLSr}{r_{obs}^\star}
\newcommand{\DMLSrad}[1]{\rad{#1}^\star}
\newcommand{\DMLSradmag}[1]{|\DMLSrad{#1}|}
\newcommand{\DMLSC}[1]{\Cx{#1}^\star}
\newcommand{\DMLSCxFun}[3]{{\DMLSC{#1}}(#2,#3)}
\newcommand{\DMLSc}[1]{\gamma_{#1}}
\newcommand{\DMLSV}{\Vobs^\star}
\newcommand{\DMLSNA}{\overline{{\Nx{\DMLSA}}_{obs}^\star}(t)}
\newcommand{\DMLSNAb}{\overline{{\Nx{\DMLSA}}_{obs}^\star}(\DMLSt{B})}
\newcommand{\DMLSNAmax}{{\overline{{\Nx{\DMLSA}}_{max}^\star}}}
\newcommand{\DMLStmax}[1]{{\DMLSt{#1}}_{,max}}
\newcommand{\DMLSdim}{\mathcal{D}}
\newcommand{\DMLSthreshInt}{\alpha^\star}

% Communication parameters
\newcommand{\data}[1]{W\left[#1\right]}
\newcommand{\dataObs}[1]{\hat{W}\left[#1\right]}
\newcommand{\thresh}{\xi}
\newcommand{\poissBar}{\Big|_\textnormal{Poiss}}
\newcommand{\gaussBar}{\Big|_\textnormal{Gauss}}
\newcommand{\Pobs}{P_{obs}}
\newcommand{\Pobsx}[1]{P_{obs}\left(#1\right)}
\newcommand{\Pone}{P_1}
\newcommand{\Pzero}{P_0}
\newcommand{\Pe}[1]{P_e\left[#1\right]}
\newcommand{\Peavg}{\overline{P}_e}
\newcommand{\threshInterval}{\alpha}

% Functions
\newcommand{\fof}[1]{f\left(#1\right)}
\newcommand{\floor}[1]{\lfloor#1\rfloor}
\newcommand{\lam}[1]{W\left(#1\right)}
\newcommand{\EXP}[1]{\exp\left(#1\right)}
\newcommand{\ERF}[1]{\erf\left(#1\right)}
\newcommand{\SIN}[1]{\sin\left(#1\right)}
\newcommand{\SINH}[1]{\sinh\left(#1\right)}
\newcommand{\COS}[1]{\cos\left(#1\right)}
\newcommand{\COSH}[1]{\cosh\left(#1\right)}
\newcommand{\Ix}[2]{I_{#1}\!\left(#2\right)}
\newcommand{\Jx}[2]{J_{#1}\!\left(#2\right)}

% Arbitrary constants
\newcommand{\B}[1]{B_{#1}}
\newcommand{\w}{w}
\newcommand{\n}{n}

\newcommand{\new}[1]{\textbf{#1}}
\newcommand{\ISI}{ISI}
\newcommand{\PDF}{PDF}

\newtheorem{theorem}{Theorem}

\maketitle

\begin{abstract}
In this paper, we apply dimensional analysis to study a diffusive molecular
communication system that uses diffusing enzymes in the propagation environment
to mitigate intersymbol interference. The enzymes bind to information molecules
and then degrade them so that they cannot interfere with the detection of
future transmissions at the receiver. We determine when
it is accurate to assume that the concentration of information molecules
throughout the receiver is constant and equal to that expected at the center of
the receiver. We show that a lower bound on the expected number of molecules
observed at the receiver can be arbitrarily scaled over the environmental
parameters, and generalize how the accuracy of the lower bound is
qualitatively impacted by those parameters.
\end{abstract}

\section{Introduction}

\IEEEPARstart{M}{olecular} communication, where a transmitter sends
information by emitting molecules into its surrounding environment to be
carried to a receiver, is a popular candidate for
implementation in networks where communicating devices have functional
components that are on the order of nanometers in size, i.e., nanonetworks.
Such networks can take advantage of the inherent biocompatibility of using
molecules as information carriers, since living organisms already do so; see
\cite[Ch. 16]{RefWorks:588}. Advancements in nanotechnology will enable a wide
range of applications using bio-hybrid components that communicate using
molecules, as described in detail in \cite{RefWorks:540, RefWorks:608}.

Free diffusion is a simple propagation method for molecules since no
external energy is required and no new infrastructure is needed between
communicating devices. However, the data transmission rate decreases as the
receiver is placed further away from the transmitter. In addition,
communication capacity is limited by the proximity of molecules over time as
they randomly diffuse. A receiver may be unable to differentiate between the
arrival of the same type of molecule emitted at different times, i.e., it can
observe intersymbol interference (\ISI). Despite these drawbacks, many
researchers have adopted free diffusion for the design of molecular
communication networks, cf. e.g. \cite{RefWorks:607, RefWorks:512,
RefWorks:548, RefWorks:574, RefWorks:609}.

Since \ISI\; is a major bottleneck to the performance of diffusive molecular
communication, actively reducing the lingering presence of information molecules
can significantly improve capacity. For example, information molecules can be
transformed as they diffuse so that they are no longer recognized by the
receiver. Specifically, enzymes diffusing in the propagation environment can be
used to repeatedly transform information molecules because of their
selectivity and because the enzymes are not consumed by the reaction mechanism.

In \cite{RefWorks:631}, we introduced a model for analyzing diffusive
molecular communication systems that have enzymes present throughout the
entire propagation environment. By reacting with the information
molecules, the enzymes improve performance; they
reduce the ``tail'' effect created by diffusing molecules that linger near
the receiver. In \cite{RefWorks:631}, we
presented a lower bound expression for the expected
number of information molecules observed at a receiver placed some distance
from a transmitter that emits impulses of molecules.

In this paper, we provide a dimensionless model for analyzing diffusive
molecular communication systems that have diffusing enzymes. Dimensional
analysis facilitates comparison between different dimensional parameter sets
with the use of reference parameters and the creation of dimensionless
constants (please refer to \cite{RefWorks:633} for more on dimensional
analysis). Using a dimensionless model generalizes the model's
scalability. Specifically, we can verify the accuracy of our lower bound
expression by simulating small environments to save computational resources and
then arbitrarily extrapolate the results to larger environments.

Furthermore, the
dimensionless model facilitates an exact study of the applicability of the
uniform concentration assumption, where the concentration of information
molecules throughout the receiver is assumed to be constant and equal to that
expected at the center of the receiver. This assumption simplifies analysis
and is often made explicitly (as in \cite{RefWorks:631, RefWorks:512}) or
implicitly (by assuming a point receiver, as in \cite{
RefWorks:548, RefWorks:574, RefWorks:609}). However, analysis of the accuracy
of this assumption has not yet been performed in the molecular communications
literature. Assessing the assumption's validity will instill confidence in its
continued use.

Dimensional analysis was applied in this paper to make the following
contributions:
\begin{enumerate}
    \item For the case of diffusion only (no active enzymes present),
    we derive the expected dimensionless number of information molecules
    observed at the receiver from emission by a point source but without the
    assumption of uniform concentration of those molecules within the receiver
    volume. We consider cubic and spherical volumes and use numerical results
    to gain insight into when the uniform concentration assumption is accurate.
    \item We show that the lower bound on the expected number of observed
    information molecules at the receiver when enzymes are present can be
    arbitrarily scaled over parameters that include the size and placement of
    the receiver, the chemical reactivity of the molecular species,
    and the number of molecules.
    \item We generalize the accuracy of the lower bound expression on the
    expected number of information molecules at the receiver when enzymes are
    present. The qualitative impact of the environmental parameters on the
    accuracy is considered.
\end{enumerate}

The rest of this paper is organized as follows. In Section~\ref{sec_model},
we describe the dimensionless model for transmission between a single
transmitter and receiver. The exact expected number of information molecules for
the case of diffusion only, where we do not apply the uniform concentration
assumption, is derived in Section~\ref{sec_diff}. In Section~\ref{sec_enz}, we
show the scalability and accuracy of the lower bound expression on the
number of molecules at the receiver when enzymes are present. Numerical and
simulation results are presented in Section~\ref{sec_results}. We make our
conclusions in Section~\ref{sec_concl}.

\section{Physical Model}
\label{sec_model}

In this section, we briefly describe our physical model as
we introduced in \cite{RefWorks:631} before we define reference parameters
to translate the model into dimensionless form.

\subsection{Dimensional Form}

The transmitter is fixed at the origin of an unbounded
3-dimensional aqueous environment. The receiver is an observer with a fixed
volume of size $\Vobs$, centered at location $\{\x_0,\y_0,\z_0\}$ where
$\rad{0}$ is the vector from the origin to $\{\x_0,\y_0,\z_0\}$, and of
arbitrary shape.

We are interested in three mobile species: $\A$ molecules, $\En$ molecules, and
$\EA$ molecules. $\A$ molecules are the information molecules that are released
by the transmitter. These molecules have a negligible natural degradation rate
but they are able to act as substrates with enzyme $\En$ molecules.
We apply the Michaelis-Menten reaction mechanism (a common mechanism for
enzymes; see \cite{RefWorks:585}) to the $\A$ and $\En$ molecules:
\begin{align}
\label{k1_mechanism}
\En + \A &\xrightarrow{\kth{1}} \EA, \\
\label{kminus1_mechanism}
\EA &\xrightarrow{\kth{-1}} \En + \A, \\
\label{k2_mechanism}
\EA &\xrightarrow{\kth{2}} \En + \AP,
\end{align}
where $\EA$ is the intermediate formed by the binding of an $\A$ molecule to an
enzyme molecule, $\AP$ is the degraded $\A$ molecule that is not recognized by
the receiver (so we can ignore them once created),
and $\kth{1}$, $\kth{-1}$, and $\kth{2}$ are the reaction rates
for the reactions as shown with units $\molecule^{-1}\metre^3\,\second^{-1}$,
$\second^{-1}$, and $\second^{-1}$, respectively. Reaction (\ref{k2_mechanism})
degrades $\A$ molecules irreversibly while the enzymes are
released intact, enabling the latter to participate in future reactions.

The number of molecules of species $\X$ is given by $\Nx{\X}$, and its
concentration at the point defined by vector $\rad{}$ and at time $t$ in
$\molecule\cdot\metre^{-3}$ is $\CxFun{\X}{\rad{}}{t}$.
For compactness, we will sometimes write $\CxFun{\X}{\rad{}}{t} = \Cx{\X}$.
We assume that every molecule of each species $\X$ diffuses independently of all
other molecules with diffusion constant $\Dx{\X}$ defined as \cite[Eq.
4.16]{RefWorks:587}
\begin{equation}
\label{JUN12_60}
\Dx{\X} = \frac{\bolt\temp}{6\pi \visc \Ri{\X}},
\end{equation}
where $\bolt$ is the Boltzmann constant ($\bolt = 1.38 \times 10^{-23}$ J/K),
$\temp$ is the temperature in kelvin, $\visc$ is the viscosity of the
medium in which the particle is diffusing ($\visc \approx
10^{-3}\,\textnormal{kg}\cdot\metre^{-1}\second^{-1}$ for water at
$25\,^{\circ}\mathrm{C}$), and $\Ri{\X}$ is the molecule radius. Thus, the units
for $\Dx{\X}$ are $\metre^2/\second$.

The transmitter communicates by emitting impulses of
$\A$ molecules, where the number of molecules emitted is $\Nx{\A}$ (not to be
confused with Avogadro's Number). The receiver counts the
number of unbound $\A$ molecules that are within the receiver volume but has no
influence on any of the reaction or diffusion processes. Using a passive
receiver without specifying a detection mechanism enables us to focus on the
propagation behaviour. $\Ne$ $\En$
molecules are initially randomly (uniformly) distributed throughout a finite
cubic volume $\Ve$ that includes both the transmitter and receiver with the
transmitter at the center. $\Ve$ is impermeable to $\En$ molecules (so that we
can simulate using a finite number of $\En$ molecules) but not $\A$ molecules
($\EA$ molecules decompose to their constituents if they hit the boundary).
Thus, the \emph{total} concentration of the free and bound enzyme in
$\Ve$, $\Cx{\Etot}$, is constant and equal to $\Ne/\Ve$. $\Ve$ is sufficiently
large to assume in analysis that it is infinite in size.

\subsection{Dimensionless Form}

For dimensional analysis we define reference variables; please refer to
\cite{RefWorks:633} for more on dimensional analysis. We define reference
concentrations in $\molecule\cdot\metre^{-3}$: $\Cx{0}$ for species $\A$,
$\Cx{\Etot}$ for $\En$, and $\kth{1}\Cx{\Etot}\Cx{0}/(\kth{-1}+\kth{2})$ for
$\EA$ (which is the maximum $\EA$ concentration for the Michaelis-Menten
mechanism in a spatially homogenous environment; see \cite{RefWorks:585}).
We define reference distance
$\dist$ in $\metre$, and we let the reference number of molecules be equal to
$\Nx{\A}$ molecules (i.e., emissions from the transmitter release one
dimensionless molecule). We then define dimensionless concentrations as
\begin{equation}
\label{AUG12_43_conc}
\DMLSC{\DMLSA} = \frac{\Cx{\A}}{\Cx{0}}, \quad
\DMLSC{\en} = \frac{\Cx{\En}}{\Cx{\Etot}}, \quad
\DMLSC{\ea} = \frac{\Cx{\EA}(\kth{-1}+\kth{2})}{\kth{1}\Cx{\Etot}\Cx{0}},
\end{equation}
for species $\A$, $\En$, and $\EA$, respectively. Similarly, dimensionless times
are defined as
\begin{equation}
\label{AUG12_43_time}
\DMLSt{\DMLSA} = \frac{\Dx{\A}t}{\dist^2}, \quad
\DMLSt{\en} = \frac{\Dx{\En}t}{\dist^2}, \quad
\DMLSt{\ea} = \frac{\Dx{\EA}t}{\dist^2},
\end{equation}
and dimensionless coordinates along the three axes are
\begin{equation}
\label{AUG12_43_coor}
\DMLSx = \frac{\x}{\dist}, \quad
\DMLSy = \frac{\y}{\dist}, \quad
\DMLSz = \frac{\z}{\dist}.
\end{equation}

Fick's Second Law, which describes the motion of particles undergoing
independent diffusion (see \cite[Ch. 4]{RefWorks:587}), can be written in
dimensionless form for species $\X$ as
\begin{equation}
\label{AUG12_46_diff}
\pbypx{\DMLSC{s}}{\DMLSt{s}} = \nabla^2\DMLSC{s},
\end{equation}
where
\begin{equation}
\pbypx{\DMLSC{s}}{\DMLSt{s}} = \pbypx{\Cx{\X}}{t}\frac{\dist^2}{\Dx{\X}\Cx{0}},
\quad
\label{AUG12_45}
\nabla^2\DMLSC{s} = \frac{\dist^2}{\Cx{0}}\nabla^2\Cx{\X}.
\end{equation}

By applying the principles of chemical kinetics (see \cite[Ch.
9]{RefWorks:585}) to the Michaelis-Menten mechanism in
(\ref{k1_mechanism})-(\ref{k2_mechanism}), the
reaction-diffusion partial differential equations are
\begin{align}
\label{AUG12_39}
\pbypx{\Cx{\A}}{t} = &\; \Da\nabla^2\Cx{\A}\! -\kth{1}\Cx{\A}\Cx{\En}\! +
\kth{-1}\Cx{\EA}, \\
\label{AUG12_40}
\pbypx{\Cx{\En}}{t} = &\; \De\nabla^2\Cx{\En}\! -\kth{1}\Cx{\A}\Cx{\En}\! +
\kth{-1}\Cx{\EA}\! + \kth{2}\Cx{\EA},\\
\label{AUG12_41}
\pbypx{\Cx{\EA}}{t} = &\; \Da\nabla^2\Cx{\EA}\! +\!\kth{1}\Cx{\A}\Cx{\En}\! -
\!\kth{-1}\Cx{\EA}\! - \!\kth{2}\Cx{\EA},
\end{align}
and they can now be re-written in dimensionless form as
\begin{align}
\label{AUG12_46}
\pbypx{\DMLSC{\DMLSA}}{\DMLSt{\DMLSA}} = &\; \nabla^2\DMLSC{\DMLSA} -
\DMLSc{1a}\DMLSC{\en}\DMLSC{\DMLSA} + \DMLSc{1a}\DMLSc{2a}\DMLSC{\ea}, \\
\label{AUG12_47}
\pbypx{\DMLSC{\en}}{\DMLSt{\en}} = &\; \nabla^2\DMLSC{\en} -
\DMLSc{\en}\DMLSC{\en}\DMLSC{\DMLSA} + \DMLSc{\en}\DMLSC{\ea},\\
\label{AUG12_48}
\pbypx{\DMLSC{\ea}}{\DMLSt{\ea}} = &\; \nabla^2\DMLSC{\ea} +
\DMLSc{\ea}\DMLSC{\en}\DMLSC{\DMLSA} - \DMLSc{\ea}\DMLSC{\ea},
\end{align}
where
\begin{align}
\label{AUG12_49}
\DMLSc{1a} = &\; \dist^2\kth{1}\Cx{\Etot}/\Dx{\A}, \quad
\DMLSc{2a} = \kth{-1}/\left(\kth{-1} + \kth{2}\right),\\
\label{AUG12_49_3}
\DMLSc{\en} = &\; \dist^2\kth{1}\Cx{0}/\Dx{\En}, \quad
\DMLSc{\ea} = \dist^2\left(\kth{-1} + \kth{2}\right)/\Dx{\EA},
\end{align}
are dimensionless constants.
Generally, two system model parameter sets with all matching dimensionless
constants are dimensionally homologous and will have the same dimensionless
solutions; see \cite{RefWorks:633}.

\section{Observations at the Receiver - Diffusion Only}
\label{sec_diff}

In this section, we derive the expected number of $\A$ molecules counted within
$\Vobs$ at time $t$ for the cases of rectangular and spherical
$\Vobs$, given that the transmitter emits $\Nx{\A}$ molecules from the origin at
$t = 0$ and there are no active enzymes.
The local point concentration of $\A$ molecules at the point defined by vector
$\rad{}$ and at time $t > 0$ can be written in dimensionless form as \cite[Eq.
4.28]{RefWorks:587}
\begin{equation}
\label{APR12_22}
\DMLSCxFun{\DMLSA}{\DMLSrad{\DMLSA}}{\DMLSt{\DMLSA}} = \frac{1}{(4\pi
\DMLSt{\DMLSA})^{3/2}}\EXP{\frac{-\DMLSradmag{\DMLSA}^2}{4 \DMLSt{\DMLSA}}},
\end{equation}
where $\DMLSrad{\DMLSA} = \rad{}/\dist$ is the vector defining
dimensionless point $\{\DMLSx,\DMLSy,\DMLSz\}$ and we recall
that one dimensionless molecule is emitted. Let $\DMLSV$ be the
dimensionless receiver volume, where each spatial dimension is scaled by
$\dist$, and let $\DMLSNA$ be the dimensionless expected (i.e., mean) number of
observed $\A$ molecules. Thus, we need to solve
\begin{equation}
\label{AUG12_59_int}
\DMLSNA = \int\limits_{\DMLSV}
\DMLSCxFun{\DMLSA}{\DMLSrad{\DMLSA}}{\DMLSt{\DMLSA}}d\DMLSV.
\end{equation}

When applying the uniform concentration assumption, as we did in dimensional
form in \cite{RefWorks:631}, we use
$\DMLSCxFun{\DMLSA}{\DMLSrad{\DMLSA}}{\DMLSt{\DMLSA}} =
\DMLSCxFun{\DMLSA}{\DMLSrad{\DMLSA,0}}{\DMLSt{\DMLSA}}
\;\forall\,\DMLSrad{\DMLSA}$, where $\DMLSrad{\DMLSA,0} = \rad{0}/\dist$. Then,
\begin{equation}
\label{AUG12_59}
\DMLSNA = \DMLSCxFun{\DMLSA}{\DMLSrad{\DMLSA,0}}{\DMLSt{\DMLSA}}\DMLSV,
\end{equation}
since $\DMLSCxFun{\DMLSA}{\DMLSrad{\DMLSA,0}}{\DMLSt{\DMLSA}}$ does not vary over
$\DMLSV$.

\subsection{Rectangular Volume}

Let us first consider $\DMLSV$ as a rectangular prism defined by
$\DMLSx_i \le \DMLSx \le \DMLSx_f$ and analogously along $\DMLSy$ and $\DMLSz$.
We present the following theorem:

\begin{theorem}[$\DMLSNA$ for Rectangular $\DMLSV$]
The expected number of $\A$ molecules counted within rectangular $\DMLSV$ when
one dimensionless molecule is released from the origin at $\DMLSt{\DMLSA} = 0$
is given by
\begin{align}
\label{APR12_38_DMLS_int}
\DMLSNA\! = &\; \int_{\DMLSx_i}^{\DMLSx_f}\!
\int_{\DMLSy_i}^{\DMLSy_f}\!
\int_{\DMLSz_i}^{\DMLSz_f}
\DMLSCxFun{\DMLSA}{\DMLSrad{\DMLSA}}{\DMLSt{\DMLSA}}d\DMLSz d\DMLSy d\DMLSx
\\
\label{APR12_38_DMLS}
= &\; \frac{1}{8}\!\!\prod_{\DMLSdim\in\{\DMLSx\!,\DMLSy\!,\DMLSz\!\}}\!\!\!
\left(\!\!\ERF{\frac{\DMLSdim_f}{2\sqrt{\DMLSt{\DMLSA}}}}\!-
\ERF{\frac{\DMLSdim_i}{2\sqrt{\DMLSt{\DMLSA}}}}\!\!\right)\!\!,
\end{align}
where $\DMLSdim\in\{\DMLSx,\DMLSy,\DMLSz\}$ represents one of the
dimensionless spatial variables, and the error function, $\ERF{\cdot}$, is an
odd function defined as
\cite[p. 406]{RefWorks:414}
\begin{equation}
\label{APR12_32}
\ERF{\w} = \frac{2}{\sqrt{\pi}}\int_0^\w \EXP{-\DMLSdim^2}d\DMLSdim.
\end{equation}
\end{theorem}
\begin{IEEEproof}
Using the substitution $v = \sqrt{b}\DMLSdim$, it is straightforward to show
that
\begin{equation}
\label{APR12_37}
\int_{\DMLSdim_1}^{\DMLSdim_2}\!\EXP{-b\DMLSdim^2}\!d\DMLSdim =
\frac{1}{2}\sqrt{\frac{\pi}{b}} \left(\!\ERF{\sqrt{b}\DMLSdim_2}-
\ERF{\sqrt{b}\DMLSdim_1}\!\right)\!.
\end{equation}

We note that $\DMLSradmag{\DMLSA}^2 = {\DMLSx}^2 + {\DMLSy}^2 + {\DMLSz}^2$, so
(\ref{APR12_38_DMLS_int}) can be separated into three independent integrals.
Using $b = \frac{1}{4\DMLSt{\DMLSA}}$, (\ref{APR12_38_DMLS}) follows.
\end{IEEEproof}

\subsection{Spherical Volume}
Next, we consider spherical $\DMLSV$. In order to solve
(\ref{AUG12_59_int}), we adjust our frame of reference so that the transmitter
is emitting from the point defined by vector $\rad{\DMLSx}$ at
$\{\DMLSx_0,0,0\}$, and the receiver volume $\DMLSV$ with dimensionless radius
$\DMLSr$ is centered at the origin. The point defined by
vector $\DMLSrad{\DMLSA}$ at $ \{\DMLSx,\DMLSy,\DMLSz\}$ is still an arbitrary
point in $\DMLSV$. The distance from the source to $\DMLSrad{\DMLSA}$ is then
$|\DMLSrad{\DMLSA} -\rad{\DMLSx}| = \sqrt{\DMLSradmag{\DMLSA}^2 +
\radmag{\DMLSx}^2 - 2\DMLSradmag{\DMLSA}\radmag{\DMLSx}\cos\phi\sin\theta}$,
where $\phi = \tan^{-1}\left(\DMLSy/\DMLSx\right)$, $\theta =
\cos^{-1}\left(\DMLSz/\DMLSradmag{\DMLSA}\right)$, and $\radmag{\DMLSx} =
\DMLSx_0$. Thus, we need to solve
\begin{equation}
\label{APR12_42_DMLS}
\DMLSNA = \int\limits_0^{\DMLSr}
\int\limits_{0}^{2\pi}
\int\limits_{0}^{\pi}
\DMLSCxFun{\DMLSA}{\DMLSrad{\DMLSA}-\rad{\DMLSx}}
{\DMLSt{\DMLSA}}\DMLSradmag{\DMLSA}^2\sin\theta
d\theta d\phi d\DMLSradmag{\DMLSA},
\end{equation}
where
\begin{align}
\label{APR12_22_DMLS_sph}
\DMLSCxFun{\DMLSA}{\DMLSrad{\DMLSA}-\rad{\DMLSx}}{\DMLSt{\DMLSA}} =
&\;(4\pi \DMLSt{\DMLSA})^{-\frac{3}{2}} \EXP{\frac{-\DMLSradmag{\DMLSA}^2
- \radmag{\DMLSx}^2}{4
\DMLSt{\DMLSA}}} \nonumber \\
& \times \EXP{\frac{\DMLSradmag{\DMLSA}\radmag{\DMLSx}
\cos\phi\sin\theta}{2 \DMLSt{\DMLSA}}}.
\end{align}

We now present the following theorem:

\begin{theorem}[$\DMLSNA$ for Spherical $\DMLSV$]
\label{theorem_spherical}
The expected number of $\A$ molecules counted within spherical $\DMLSV$ when one
dimensionless molecule is released from $\rad{\DMLSx}$ at $\DMLSt{\DMLSA} = 0$
is given by
\begin{align}
\DMLSNA = &\; \frac{1}{2}\left[\ERF{\frac{\DMLSr\!-
\radmag{\DMLSx}}{2\sqrt{\DMLSt{\DMLSA}}}} +
\ERF{\frac{\DMLSr\!+\radmag{\DMLSx}}{2\sqrt{\DMLSt{\DMLSA}}}}\right] \nonumber
\\
& +
\frac{1}{\radmag{\DMLSx}}\sqrt{\frac{\DMLSt{\DMLSA}}{\pi}}
\bigg[\EXP{-\frac{(\radmag{\DMLSx}+\DMLSr)^2}{4\DMLSt{\DMLSA}}} \nonumber \\
& - \EXP{-\frac{(\radmag{\DMLSx}-\DMLSr)^2}{4\DMLSt{\DMLSA}}}\bigg].
\label{JUN12_57_DMLS}
\end{align}
\end{theorem}
\begin{IEEEproof}
Please refer to the Appendix.
\end{IEEEproof}

Due to symmetry, (\ref{JUN12_57_DMLS}) applies to the general case of any
spherical $\DMLSV$ whose center is at a distance equal to $\radmag{\DMLSx}$ from
a point transmitter. Interestingly, also due to symmetry, (\ref{JUN12_57_DMLS})
is analogous to the concentration at a \emph{point receiver} due to a
\emph{spherical transmitter} releasing a uniform distribution of $\A$
molecules, as given in \cite[Eq. 3.8]{RefWorks:586}. A comparison between the
expected number of observed molecules in spherical receivers with the
number expected when we apply the uniform concentration assumption
is made in Section~\ref{sec_results}.

\section{Observations at the Receiver - Enzymes Present}
\label{sec_enz}

In this section, we consider the presence of active enzymes. We have no
analytical solution to the system of equations defined by
(\ref{AUG12_46})-(\ref{AUG12_48}). Following our reasoning applied
in \cite{RefWorks:631}, where we used the uniform concentration
assumption and assumed that $\kth{-1}\Cx{\EA} \to 0$
and $\Cx{\En} \approx \Cx{\Etot}$ (where $\Cx{\Etot}$ is the
dimensional total enzyme concentration),
the lower bound on the expected concentration of
molecules when enzymes are present can be written in dimensionless form as
\begin{equation}
\label{AUG12_58}
\DMLSC{\DMLSA} \ge \frac{1}{(4\pi \DMLSt{\DMLSA})^{3/2}}
\EXP{-\frac{\dist^2\kth{1}\Cx{\Etot}}{\Da}\DMLSt{\DMLSA}
-\frac{\DMLSradmag{\DMLSA,0}^2}{4\DMLSt{\DMLSA}}}.
\end{equation}

An explicit discussion of the accuracy of (\ref{AUG12_58}) would require a
non-bounding expression for the expected number of molecules. We
derived a lower bound, therefore
we make a comparison between (\ref{AUG12_46})-(\ref{AUG12_48}) and the
system of partial differential equations that has (\ref{AUG12_58}) as its
\emph{exact} solution in order to make qualitative statements about the accuracy
of (\ref{AUG12_58}). The latter system is
\begin{align}
\label{AUG12_55}
\pbypx{\DMLSC{\DMLSA}}{\DMLSt{\DMLSA}}\bigg|_{\bound} = &\;
\nabla^2\DMLSC{\DMLSA} -
\frac{\dist^2\kth{1}\Cx{\Etot}}{\Da}\DMLSC{\DMLSA}, \\
\pbypx{\DMLSC{\en}}{\DMLSt{\en}}\bigg|_{\bound} = &\;0, \\
\label{AUG12_55_2}
\pbypx{\DMLSC{\ea}}{\DMLSt{\ea}}\bigg|_{\bound} = &\;0,
\end{align}
where $\DMLSC{\en} = 1$ is constant and $\DMLSC{\ea} = 0$.
Unlike (\ref{AUG12_46})-(\ref{AUG12_48}), this system of equations has a single
dimensionless constant, $\DMLSc{1a_{\bound}}$,
\begin{equation}
\label{AUG12_57}
\DMLSc{1a_{\bound}} = \dist^2\kth{1}\Cx{\Etot}/\Da =
\DMLSc{1a}.
\end{equation}

The accuracy of (\ref{AUG12_58}) in dimensionless form can then be written as
the differences between (\ref{AUG12_46})-(\ref{AUG12_48}) and
(\ref{AUG12_55})-(\ref{AUG12_55_2}), i.e.,
\begin{align}
\label{AUG12_56}
\pbypx{\DMLSC{\DMLSA}}{\DMLSt{\DMLSA}} -
\pbypx{\DMLSC{\DMLSA}}{\DMLSt{\DMLSA}}\bigg|_{\bound} = &\;
\frac{\dist^2}{\Da}\Cx{\EA}\left(\frac{\kth{-1}}{\Cx{0}} +
\kth{1}\DMLSC{\DMLSA}\right),
\\
\label{AUG12_56_2}
\pbypx{\DMLSC{\en}}{\DMLSt{\en}} -
\pbypx{\DMLSC{\en}}{\DMLSt{\en}}\bigg|_{\bound} =
&\; \nabla^2\DMLSC{\en} -
\DMLSc{\en}\DMLSC{\en}\DMLSC{\DMLSA} + \DMLSc{\en}\DMLSC{\ea},
\\
\label{AUG12_56_3}
\pbypx{\DMLSC{\ea}}{\DMLSt{\ea}} -
\pbypx{\DMLSC{\ea}}{\DMLSt{\ea}}\bigg|_{\bound} =
&\; \nabla^2\DMLSC{\ea} +
\DMLSc{\ea}\DMLSC{\en}\DMLSC{\DMLSA} - \DMLSc{\ea}\DMLSC{\ea},
\end{align}
where we used $\Cx{\Etot} = \Cx{\En} + \Cx{\EA}$, even though $\Cx{\Etot}$ was
not defined as a function of time and space. Of course, (\ref{AUG12_56_2}) and
(\ref{AUG12_56_3}) are identical to (\ref{AUG12_47}) and (\ref{AUG12_48}),
respectively. Eqs. (\ref{AUG12_56})-(\ref{AUG12_56_3}) show what we
lose when we consider the lower bound (\ref{AUG12_58}). Particularly, we lose
all of the reaction-diffusion dynamics of the $\En$ and $\EA$ molecules.
Eqs. (\ref{AUG12_56_2}) and (\ref{AUG12_56_3}) have both positive and negative
terms, so we focus on (\ref{AUG12_56}) to make comments about the
accuracy of (\ref{AUG12_58}). This is acceptable because we are ultimately most
interested in the dynamics of the $\A$ molecules.

From (\ref{AUG12_56}), we can immediately claim:
\begin{enumerate}
    \item A higher $\kth{1}$, $\kth{-1}$, or an increase in $\Nx{\A}$ or
    $\Nx{\En}$ in an otherwise unchanged system will
    decrease the accuracy of (\ref{AUG12_58}), because having more $\A$ or $\En$
    molecules indirectly increases $\Cx{\EA}$.
    \item A higher $\kth{2}$ in an otherwise
    unchanged system will increase the accuracy of (\ref{AUG12_58}), because
    there will be fewer $\EA$ molecules throughout the system. This observation
    is also intuitive given that $\kth{2} \to \infty$ leads to our assumption
    that $\kth{-1}\Cx{\EA} \to 0$.
    \item The impact of scaling $\dist$ and $\Dx{\A}$ is non-trivial because
    both $\DMLSC{\DMLSA}$ and $\Cx{\EA}$ are functions of location.
    A larger $\dist$ or smaller $\Dx{\A}$ might decrease the
    accuracy of (\ref{AUG12_58}), but increasing $\dist$ or decreasing
    $\Dx{\A}$ would also mean that the observed $\DMLSC{\DMLSA}$ and $\Cx{\EA}$
    decrease, which we have already said would increase the accuracy.
\end{enumerate}

The cumulative impact on the accuracy of (\ref{AUG12_58}) when varying the
environmental parameters is better appreciated with simulations, as we present
in Section~\ref{sec_results}.

\section{Numerical and Simulation Results}
\label{sec_results}

\subsection{Uniform Concentration Assumption}

\begin{figure}[!tb]
\centering
\includegraphics[width=\linewidth]{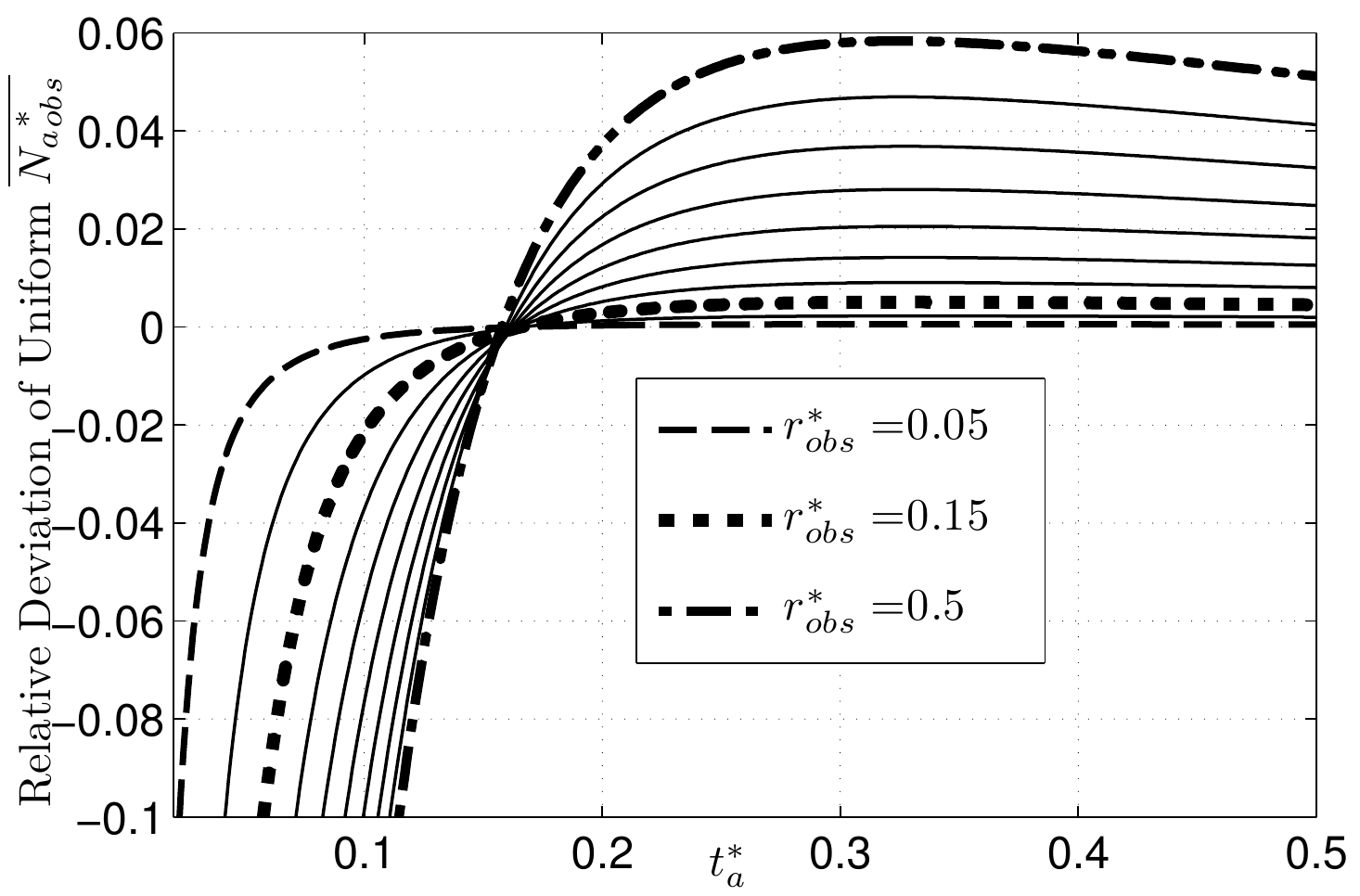}
\caption{The relative deviation in $\DMLSNA$ from the true value
(\ref{JUN12_57_DMLS}) at a spherical receiver when the uniform concentration
assumption (\ref{AUG12_59}) is applied without active enzymes. The deviation is
evaluated as ((\ref{AUG12_59})-(\ref{JUN12_57_DMLS}))/(\ref{JUN12_57_DMLS}).
The spherical receiver radius
$\DMLSr$ is increased in increments of $0.05$, where a larger $\DMLSr$ results
in a larger deviation.}
\label{uniform_test}
\end{figure}

We first present a numerical test of the uniform concentration assumption. We
set $\dist = \rad{0}$; we are only interested in $\DMLSr < 1$.
Small values of $\DMLSr$ correspond to a smaller receiver or the receiver
placed further from the transmitter.
In Fig.~\ref{uniform_test}, we show how much $\DMLSNA$ at a spherical
receiver of radius $\DMLSr$ deviates from the true value over time when we
assume that the concentration throughout $\DMLSV$ is uniform. A similar test
was also performed for the deviation at a cube, and the results
were visually indistinguishable to Fig.~\ref{uniform_test} so they are omitted
here due to space (though they show that the exact shape of $\DMLSV$ is not
important).

We see in Fig.~\ref{uniform_test} that the deviation from the true value of
$\DMLSNA$ increases with $\DMLSr$. There is
significant deviation for any $\DMLSr$ when $\DMLSt{\DMLSA}$ is
sufficiently small. $\DMLSNA$ is underestimated for all $\DMLSr$ until
$\DMLSt{\DMLSA} \approx 0.16$ and then it is overestimated, but the
deviation tends to $0$ as $\DMLSt{\DMLSA} \to \infty$.
This transition is intuitive; molecules tend to diffuse to the edge of $\DMLSV$
before they reach the center.
However, the centre of $\DMLSV$ is closer to the transmitter than most of
$\DMLSV$, leading to the eventual overestimate of $\DMLSNA$.

The deviation in $\DMLSNA$ should be no more than a few percent.
We are generally interested in values of
$\DMLSt{\DMLSA} > 0.1$, so the initial large deviation for all values of
$\DMLSr$ is not a major concern. Small values of $\DMLSr$, i.e., $\DMLSr \le
0.15$, maintain a deviation of less than $2\,\%$ for all $\DMLSt{\DMLSA} > 0.1$.
Thus, we claim that the uniform concentration assumption is sufficient
when studying receivers whose radius is no more than 15\,\% of
the distance from the center of the receiver to the transmitter.

\subsection{Dimensionless Accuracy of Expected Number of Molecules}

We now consider two systems with different environmental parameters in order to
assess the accuracy of the lower bound expression (\ref{AUG12_58}) and highlight
the scalability of the dimensionless model.
The details of our particle-based simulation framework are described in
\cite{RefWorks:631}. We assume that each system has a viscosity of
$10^{-3}\,\textnormal{kg}\,\metre^{-1}\second^{-1}$ and temperature of
$25\,^{\circ}\mathrm{C}$. $\Ve$ is defined as a cube with side length 1\,$\mu$m,
and centered at the origin. We set simulation time step $\Delta t =
0.5\,\mu\second$.
The reference parameters are $\dist = \radmag{0} = 300/\sqrt{2}$\,nm and $\Cx{0}
= \frac{\Nx{\A}}{1\,\metre^3}$. Each system has a spherical receiver with
dimensionless radius $\DMLSr = 0.15$. The radii of the molecular species are 
$\Ri{\A} = 0.5\,$nm, $\Ri{\En} = 2.5\,$nm, and $\Ri{\EA} = 3\,$nm.
The rate constants $\kth{-1}$ and $\kth{2}$ are $2\times10^4\,\second^{-1}$ and
$2\times10^6\,\second^{-1}$, respectively. The unique system parameters are
described in Table~\ref{table_accuracy}. The number of molecules and
the size of the environments are kept deliberately
low to ease simulation time.

One can quickly
verify that the two systems are dimensionally homologous according to the
constants in (\ref{AUG12_49})-(\ref{AUG12_49_3}). They have the same
lower bound on the expected number of observed molecules as defined by
(\ref{AUG12_58}), and have the same loss in accuracy as defined by
(\ref{AUG12_56}) when we use $\kth{1}\Cx{\Etot}\Cx{0}/(\kth{-1}+\kth{2})$
for $\Cx{\EA}$ and $1$ in place of $\DMLSC{\DMLSA}$.

\begin{table}[!tb]
\centering
\caption{System parameters used for Fig.~\ref{fig_accuracy}.}
{\renewcommand{\arraystretch}{1.4}
	\begin{tabular}{|c|c|c|}
	\hline
	Parameter & System 1 & System 2\\ \hline
	$\Nx{\A}$ & $10^4$ & $2\times10^4$ \\ \hline
	$\Nx{\Etot}$ & $2\times10^5$ & $4\times10^5$ \\ \hline
	$\kth{1}$ [$\frac{\metre^3}{\molecule\cdot\second}$]& $2\times10^{-19}$ &
	$10^{-19}$ \\ \hline
	\end{tabular}
}
\label{table_accuracy}
\end{table}

\begin{figure}[!tb] \centering \includegraphics[width=\linewidth]
{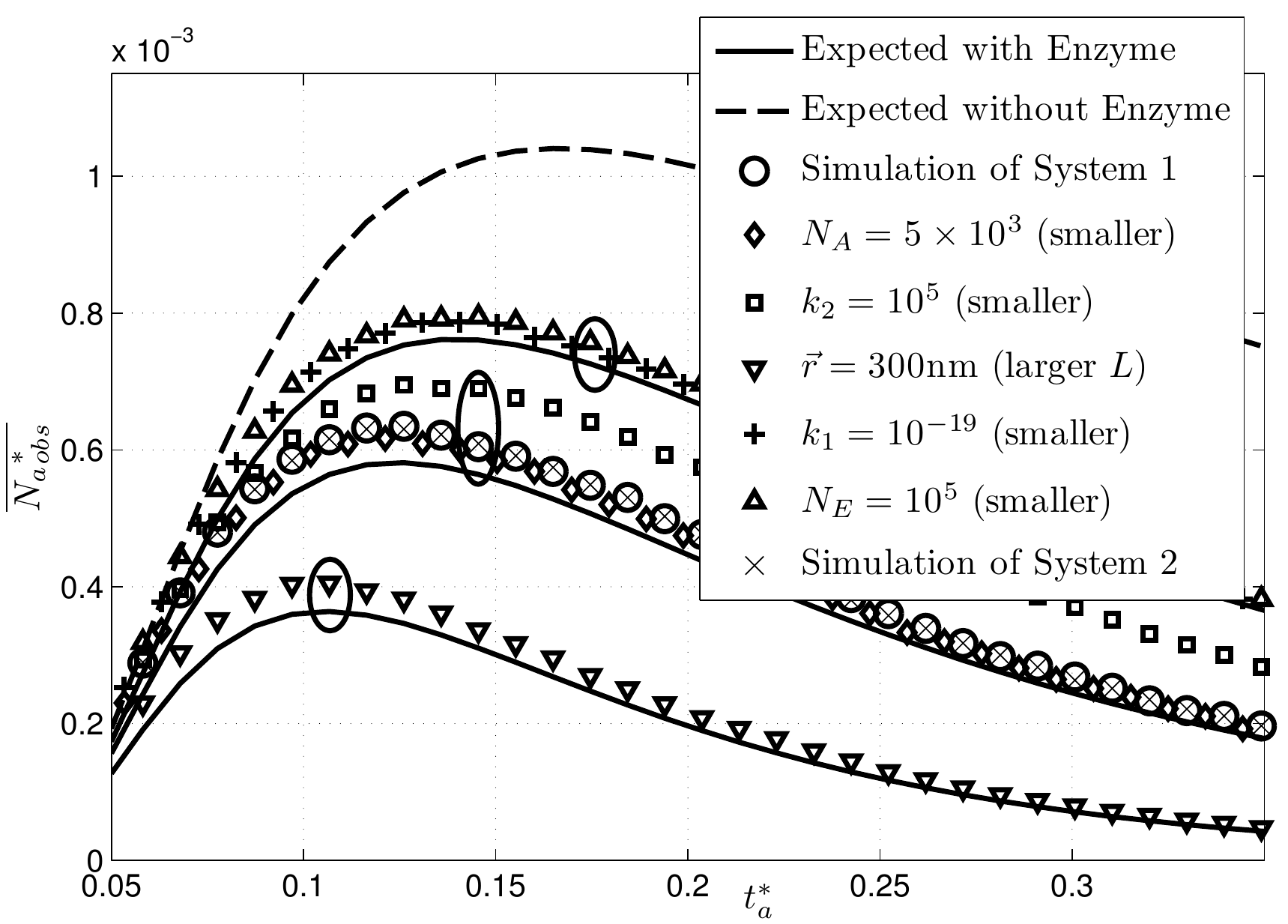}
\caption{Assessing 1) the accuracy of the lower bound on
the expected number of observed information
molecules and 2) the consistency of dimensionally
homologous systems. Systems 1 and 2 are dimensionally homologous and
the modified versions of System 1 are not, although
many share common lower bounds on $\DMLSNA$ (as grouped by ellipses).
The units for the modified rate constants are the same as those listed
in the text and in
Table~\ref{table_accuracy}. For reference, the maximum dimensional expected
number of molecules observed for Systems 1 and 2 when enzymes are present
is 5.81 and 11.63 molecules at 12.84\,$\mu$s, respectively.}
\label{fig_accuracy}
\end{figure}

In Fig.~\ref{fig_accuracy}, we compare the observed number of
dimensionless molecules using the two sets of system parameters, in addition to
variations of System 1 with a single parameter modified (as noted in the legend,
and these variations are not homologous).
The number of $\A$ molecules observed via simulation
is averaged over at least 6000
independent emissions by the transmitter at $\DMLSt{\DMLSA} = 0$. We measure
the dimensionless number of information molecules observed over dimensionless
time, in comparison to the lower bound expression (\ref{AUG12_58}) and the
expected number without enzymes in the environment as given by (\ref{AUG12_58})
for $\Cx{\Etot} = 0$ (where in both cases the concentrations given by
(\ref{AUG12_58}) are multiplied by $\DMLSV$).

We clearly see in Fig.~\ref{fig_accuracy} that the receivers in Systems 1 and 2
observe, on average, the same number of dimensionless information
molecules. The changes in accuracy for the modified versions
of System 1 are consistent with the claims made from (\ref{AUG12_56});
the accuracy of (\ref{AUG12_58}) improves with lower $\Nx{\A}$,
$\Nx{\En}$, $\kth{1}$, or higher $\kth{2}$. With an increase in $\dist$, the
accuracy appears to be worse initially and then improves more rapidly over time.
Importantly, $\DMLSNA$ as shown in most curves is much closer to the lower
bounds than to the expected value without enzymes (without enzymes, all systems
here have the same $\DMLSNA$).
This accuracy is achieved even though these systems have a high $\kth{1}$ value
(the largest possible value of $\kth{1}$ is on the order of $1.66\times
10^{-19}\,\molecule^{-1}\metre^3\,\second^{-1}$; see \cite[Ch.
10]{RefWorks:585}) and a large enzyme concentration (the lowest concentration
considered here, $166\,\mu\textnormal{M}$ when System 1 is modified so that
$\Nx{\Etot} = 10^5$, is still high for a cellular enzyme; see
\cite{RefWorks:632}).  Thus, the lower bound expression (\ref{AUG12_58}) can be
applied in future work to derive communications performance when enzymes are
added to mitigate \ISI.
Furthermore, the dimensional analysis enables us to simulate small
environments, even if they are too small for a practical implementation with
modern technology, and then scale the results to arbitrarily larger but
dimensionally homologous environments with many more molecules, thereby saving
computational resources.

\section{Conclusions}
\label{sec_concl}

In this paper, we used dimensional analysis to study a diffusive molecular
communication system where diffusing enzymes in the propagation environment
mitigate intersymbol interference. We derived the exact expected number of
information molecules observed at cubic and spherical receivers when enzymes are
not present, and showed that the uniform concentration assumption is
sufficiently accurate when the radius of the receiver is no more than
15\,\% of the distance from the transmitter to the center of the receiver. We
also showed that, when active enzymes are present, the accuracy of the lower
bound expression on the expected number of molecules can be qualitatively
described with respect to the environmental parameters.
Adopting a dimensionless model saves computational resources for
analytical verification by
simulating environments with fewer molecules.
On-going work is using the lower bound to assess the design of practical signal
detection schemes for the receiver in such an environment with derivation of the
corresponding bit error probabilities to more clearly demonstrate the
improvement in communication performance when using enzymes.

\appendix

To prove Theorem~\ref{theorem_spherical}, we first integrate
(\ref{APR12_42_DMLS}) with respect to $\phi$ by using \cite[Eq.
3.339]{RefWorks:402}
\begin{equation}
\label{JUN12_02_DMLS}
\int_0^\pi \EXP{b\cos \w}d\w = \pi\Ix{0}{b},
\end{equation}
where $\Ix{0}{b}$ is the modified zeroth order Bessel function of the first
kind. It is straightforward to show that the integral of $\EXP{b\cos \w}$ from
$\w = \pi$ to $\w = 2\pi$ is also $\pi\Ix{0}{b}$, so from (\ref{APR12_42_DMLS})
we can write
\begin{equation}
\label{JUN12_15_DMLS}
\int_0^{2\pi}\!\!\EXP{\frac{\DMLSradmag{\DMLSA}
\radmag{\DMLSx}\cos\phi\sin\theta}{2 \DMLSt{\DMLSA}}}\!d\phi =
2\pi\Ix{0}{\frac{\DMLSradmag{\DMLSA}\radmag{\DMLSx}\sin\theta}{2\DMLSt{\DMLSA}}}.
\end{equation}

Next, we integrate with respect to $\theta$ by using
\cite[Eq. 6.681.8]{RefWorks:402}
\begin{equation}
\label{JUN12_16}
\int\limits_0^\pi \!\SIN{2\mu\w}\Jx{2\nu}{2b \sin\w}d\w =
\pi\SIN{\mu\pi}\Jx{\nu-\mu}{b}\Jx{\nu+\mu}{b},
\end{equation}
where $\Jx{i}{b}$ is the $i$th order Bessel function of the first kind and
$\Ix{0}{b} = \Jx{0}{jb}$, where $j = \sqrt{-1}$. From (\ref{APR12_42_DMLS}),
(\ref{JUN12_15_DMLS}), and (\ref{JUN12_16}) we can write
\begin{multline}
\label{JUN12_17_DMLS}
\int_0^\pi \sin\theta
\Ix{0}{\frac{\DMLSradmag{\DMLSA}\radmag{\DMLSx}\sin\theta}{2\DMLSt{\DMLSA}}}d\theta = \\
\pi\Jx{-\frac{1}{2}}{\frac{j\DMLSradmag{\DMLSA}\radmag{\DMLSx}}{4\DMLSt{\DMLSA}}}
\Jx{-\frac{1}{2}}{\frac{j\DMLSradmag{\DMLSA}\radmag{\DMLSx}}{4\DMLSt{\DMLSA}}}.
\end{multline}

Using \cite[Eqs. 1.311, 1.334, 8.464.1-2]{RefWorks:402}, it is straightforward
to show that
\begin{equation}
\label{JUN12_19_23}
\Jx{-\frac{1}{2}}{jb}\Jx{\frac{1}{2}}{jb} = \frac{1}{2\pi b}
\left(\EXP{2b} - \EXP{-2b}\right),
\end{equation}
so (\ref{APR12_42_DMLS}) is now reduced to
\begin{align}
\label{JUN12_50_DMLS}
\DMLSNA = &\; \frac{1}{2\radmag{\DMLSx}\sqrt{\pi\DMLSt{\DMLSA}}}
\int_0^{\DMLSr}
\DMLSradmag{\DMLSA}\bigg[\EXP{-
\frac{(\radmag{\DMLSx}-\DMLSradmag{\DMLSA})^2}{4\DMLSt{\DMLSA}}} \nonumber \\
& - \EXP{-\frac{(\radmag{\DMLSx}+\DMLSradmag{\DMLSA})^2}
{4\DMLSt{\DMLSA}}}\bigg]d\DMLSradmag{\DMLSA},
\end{align}
which can be solved using substitution to arrive at (\ref{JUN12_57_DMLS}).
\hfill\IEEEQED

\bibliography{../references/nano_ref}

\end{document}